\documentclass[12pt,english]{article}

\usepackage{geometry}
\usepackage[titletoc,title]{appendix}

\usepackage{setspace}
\usepackage{sectsty}
\allsectionsfont{\normalsize\raggedright\centering}

\usepackage[]{tocloft}
\addtocontents{toc}{\cftpagenumbersoff{section}}
\addtocontents{toc}{\cftpagenumbersoff{subsection}}

\makeatletter
\renewcommand{\@cftmaketoctitle}{}
\makeatother

\usepackage{natbib}
\usepackage{url}
\makeatletter
\def\url@leostyle{%
    \def\UrlFont{\sf}}{\def\UrlFont{\small\ttfamily}}
\makeatother
\bibpunct
{(} 
{)} 
{,} 
{a} 
{} 
{;} 

\usepackage{authblk}

\usepackage{graphicx}
\usepackage{babel}

\usepackage{amsmath}
\usepackage{amsfonts}

\usepackage{framed}

\usepackage{xifthen}


\numberwithin{equation}{section}

\usepackage{txfonts}
\usepackage[T1]{fontenc}
\usepackage{mathrsfs}
\newcommand{\citetbjps}[2][]{\ifthenelse{\equal{#1}{}}{\citeauthor{#2} ([\citeyear{#2}])}{\citeauthor{#2} ([\citeyear{#2}], #1)}}
\newcommand{\citealtbjps}[2][]{\ifthenelse{\equal{#1}{}}{\citeauthor{#2} [\citeyear{#2}]}{\citeauthor{#2} [\citeyear{#2}], #1}}
\newcommand{\citepbjps}[2][]{\ifthenelse{\equal{#1}{}}{(\citeauthor{#2} [\citeyear{#2}])}{(\citeauthor{#2} [\citeyear{#2}], #1)}}
\newcommand{\citeyearbjps}[2][]{\ifthenelse{\equal{#1}{}}{[\citeyear{#2}]}{[\citeyear{#2}], #1}}
\newcommand{\citeyearparbjps}[2][]{\ifthenelse{\equal{#1}{}}{([\citeyear{#2}])}{([\citeyear{#2}], #1)}}
\newcommand{\citeposbjps}[2][]{\ifthenelse{\equal{#1}{}}{\citeauthor{#2}'s ([\citeyear{#2}])}{\citeauthor{#2}'s ([\citeyear{#2}], #1)}}

\begin{document}

\title{{\Large \textbf{On the Universality of Hawking Radiation}}}
\author{{\normalsize \textbf{Sean Gryb, Patricia Palacios, and Karim Th\'{e}bault}}\\\textit{\normalsize The British Journal for the Philosophy of Science} \normalsize (forthcoming) \\\normalsize\today}

\date{}

\maketitle

\thispagestyle{empty}

\begin{abstract}

A physically consistent semi-classical treatment of black holes requires universality arguments to deal with the `trans-Planckian' problem where quantum spacetime effects appear to be amplified such that they undermine the entire semi-classical modelling framework. We evaluate three families of such arguments in comparison with Wilsonian renormalization group universality arguments found in the context of condensed matter physics. Our analysis is framed by the crucial distinction between robustness and universality. Particular emphasis is placed on the quality whereby the various arguments are underpinned by `integrated' notions of robustness and universality.  Whereas the principal strength of Wilsonian universality arguments can be understood in terms of the presence of such integration, the principal weakness of all three universality arguments for Hawking radiation is its absence.
\end{abstract}

\tableofcontents

\section{Introduction}

Universality arguments have been developed in the context of black holes physics to resolve a problem in Hawking's famous prediction that black holes produce thermal radiation \citepbjps{hawking:1975}. As was recognized soon after Hawking's original paper \citepbjps{gibbons:1977}, the derivation of Hawking radiation makes essential use of a breakdown in the separation between micro- and macro-scales. In particular, due to an exponential redshift, the Hawking radiation that is detected far away from the black hole, at relatively large wavelengths and late times, is sensitive to the near horizon physics at ultra-short wavelengths, even much smaller than the Planck scale of $10^{-35}$ m.  This `trans-Planckian problem' is not just a curious side-note. Rather it implies that quantum spacetime effects can be amplified such that they undermine the semi-classical framework for modelling black holes. 

The strongest lines of response to the trans-Planckian problem that are available in the literature are all based upon `universality' arguments. What each of these sets of arguments have in common is that they aim to establish that the phenomenon of Hawking radiation is both `universal' and `robust'. Universality in this context is  insensitivity to variation of the macroscopic details that characterise the type of black hole considered (for example, stationary versus non-stationary spacetime). Robustness in this context is insensitivity to variation of the microphysical details that characterise the particular token of black hole considered (for example, different ultraviolet physics at the horizon). In this paper we consider three sets of arguments designed to establish the universality and robustness of Hawking radiation. These are based upon respectively: i) the Unruh effect and equivalence principle (\citealtbjps{Agullo:2009wt}; \citealtbjps{Wallace:2017a}); ii) horizon symmetries (\citealtbjps{Birmingham:2001qa}; \citealtbjps{PhysRevD.77.024018}; \citealtbjps{Iso:2006wa}); and iii) modified dispersion relations (\citealtbjps{unruh:2005}; \citealtbjps{himemoto:2000}; \citealtbjps{Barcelo:2008qe}). 

The structure of the highly successful `Wilsonian' universality arguments used in condensed matter physics is the starting point for our present analysis.\footnote{For philosophical discussion see (Batterman \citeyearbjps{batterman:2000,batterman:2002}; \citealtbjps{Mainwood2006}; \citealtbjps{butterfield:2011}; \citealtbjps{ruetsche:2011}; \citealtbjps{Franklin2017}; \citealtbjps{Palacios2017}; \citealtbjps{Fraser:2018}; \citealtbjps{shech:2018}; \citealtbjps{saatsi:2018}).} Our goal is to use these arguments  to establish a general framework within which other universality arguments, including those for Hawking radiation, can be evaluated. Our framework is based upon six qualities that characterize Wilsonian universality arguments: 
\begin{itemize}
\item []1. Degree of Robustness. The range of single-type token-level variation across which the invariance of the relevant phenomena can be established; 
\item [] 2. Physical Plausibility. The applicability of the robustness and universality arguments to de-idealized target systems; 
\item [] 3. Degree of Universality. The range of inter-type variation across which the invariance of the relevant phenomena can be established; 
\item [] 4. Comprehensiveness. The size of the set of observables over which the robustness argument can be applied; 
\item [] 5. Empirical Support. Experimental evidence supporting instantiation of effects in different types and/or relevant methods of approximation; 
\item [] 6. Integration. Feature whereby the theoretical bases behind the invariance found in the universality arguments and the robustness arguments are mutually consistent.
\end{itemize}  

It will be argued that whereas Wilsonian universality arguments have a high score in all these six qualities, universality arguments for Hawking radiation fail to measure up to their Wilsonian counterparts. In particular, it will be emphasized the strength of the Wilsonian arguments is predicated upon their integration: they can combine such a high degree of robustness and universality because the theoretical bases behind the invariance are mutually consistent. Furthermore, it is this combination that allows the arguments to also be highly physically plausible (Batterman \citeyearbjps{batterman:2000,batterman:2002}; \citealtbjps{batterman:2014}). The converse is found to hold in the Hawking case: the universality arguments that do well on degree of robustness do poorly on degree of universality and physical plausibility, and arguments that do well on degree of universality do poorly on degree of robustness. We argue that the lack of integration found in these arguments is a plausible reason for the existence of these trade-offs between the relevant subset of qualities. This negative conclusion provides a possible basis for future scientific research aimed at developing integrated universality arguments for Hawking radiation.

\section{Hawking Radiation and the Trans-Planckian Problem}
\label{sec:Hawking}

\subsection{What is Hawking Radiation?}
In quantum field theory on curved spacetime the ground state of a free quantum field can evolve into a state with excitations when the geometry of the background spacetime is non-trivial. In an asymptotically flat spacetime, this means that there can be a flux of `out-particles' at future-null infinity $\mathscr I^+$ even if there is no flux of `in-particles' at past-null infinity $\mathscr I^-$ \cite[\S5]{jacobson:2005}. Most famously when the spacetime in question features a black hole, this flux can have a characteristic thermal form that depends only upon intrinsic properties of the black hole spacetime \citepbjps{KAY199149}. This thermalized flux is the Hawking radiation of a black hole.

\begin{figure}[h]
	\begin{center}
		{ \includegraphics[width=0.55\textwidth]{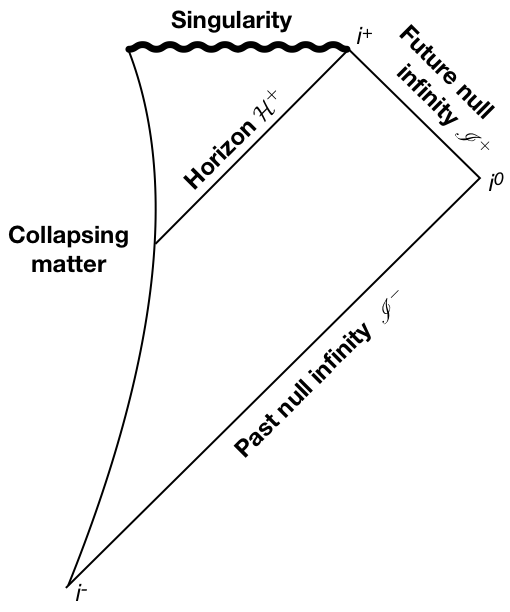}
  }
	\end{center}
	\caption{\label{fig:bh diagram} The conformal diagram of a spacetime external to a spherically symmetric distribution of collapsing matter.}
\end{figure}

The original treatment of \citealtbjps{hawking:1975} proceeds as follows. Model an astrophysical black hole spacetime as the formation of a Schwarzschild black hole from the collapse of a matter shell.\footnote{See \citealtbjps{dewitt1975quantum} for the Kerr case.} The conformal diagram for the spacetime exterior to the collapsing matter is given in Figure~\ref{fig:bh diagram}. Within the spacetime assume that there is a free quantized Klein--Gordon field $\hat\phi$. Further assume that there is no back-reaction between the quantum field and the background spacetime. The asymptotic flatness of the spacetime before and after the collapse selects unique ground state vacua and Fock representations for the Klein--Gordon field on $\mathscr I^\pm$. 
Assuming that there is no incoming radiation from $\mathscr{I}^{-}$, Hawking used a frame-independent procedure to show that the evolution of such a state would appear, to first order, as a thermal state at the Hawking temperature $T_H=\frac{\hbar\kappa}{2\pi}$ for an observer near $\mathscr I^+$ at asymptotically late times---that is, near future time-like infinity $i^+$---where $\kappa$ is the surface gravity of the black hole horizon.

Various proposals have been made to provide a local physical mechanism for the production of Hawking radiation. The different proposals vary significantly in terms of where and how the thermal radiation is produced and are largely mutually inconsistent. The most significant possible mechanisms include:  splitting of entangled modes as the horizon forms (\citealtbjps{unruh:1977}; \citealtbjps{gibbons:1977}); tidal forces pulling apart virtual particle--anti-particle pairs (\citealtbjps{hawking:1979}; \citealtbjps{adler:2001}; \citealtbjps{dey:2017}); entangled radiation quantum tunnelling through the horizon \citepbjps{parikh:2000}; the effects of non-stationarity of the background metric field (\citealtbjps{fredenhagen:1990}; \citealtbjps{jacobson:2005}); and anomaly cancellation \citepbjps{PhysRevD.77.024018}. The formal rigour of these proposals varies greatly, and none is entirely satisfactory from a physical perspective. 

Ideally, what we want to find is a relationship between the way in which the gravitational degrees of freedom interact with $\hat\phi$ during the evaporation process and the production of the radiation itself. An explicit such demonstration is currently lacking for most, if not all, of the putative mechanisms \citepbjps{Curiel:2018}. At the very least, to be physically plausible, a mechanism for Hawking radiation must be explicitly demonstrated to be generalisable from the highly idealized cases of eternal black holes to physically realistic astrophysical models in which time translation symmetry is broken by the collapse phase leading up to the formation of the horizon \citepbjps{hollands:2015}.

Notwithstanding this lack of unique physically plausible mechanism or region of origin associated with Hawking radiation, it is certainly significant that the formal expression for Hawking flux has proved `remarkably robust' under the inclusion of various complicating factors (\citealtbjps{leonhardt:2008}; \citealtbjps{thompson:2008}) and formal clarifications \citepbjps{fredenhagen:1990}. As noted by \citealtbjps{Wallace:2017a}, such consistency between various theoretical derivations strongly suggests that Hawking radiation really is a consequence of semi-classical gravity and not simply an artefact of a particular (potentially flawed) argument. That said, robustness of an effect between different theoretical models offers little comfort if there is a problematic feature that all the various derivations share.

\subsection{Red-Shift and Robustness}  

An important insight into the physics behind Hawking radiation, which is independent of the mechanism of production, can be obtained by considering the possible equilibrium states of the quantum field long after the collapse process has taken place. By this time we can assume that the geometry will have effectively reached Schwarzschild form (assuming zero angular momentum). Since this geometry is not maximally symmetric, there is no unique ground state singled-out by the global symmetries. A variety of physically motivated principles  can be used to resolve this underdetermination. These principles lead to different choices of privileged observers, who are required to see the quantum field in a ground state. One choice, the `Boulware vacuum' \citepbjps{boulware1975quantum}, is a ground state as seen by the static observers of the exterior Schwarzschild geometry. In this state no fluxes are observed and there is thus no Hawking radiation. The Boulware vacuum is irregular on the horizon and thus unphysical in this context. Two vacua that are regular on the horizon are the `Hartle--Hawking vacuum' \citepbjps{hartle1976path} and the `Unruh vacuum' (\citealtbjps{unruh1976notes}; \citealtbjps{dappiaggi2011rigorous}). 

The Hartle--Hawking vacuum is a ground state as seen by in- and out-going null observers of the entire maximally-extended Schwarzschild geometry. When restricted to the exterior geometry, this state appears as a thermal state near future time-like infinity $i^+$ for modes out-going from near the horizon \citepbjps{KAY199149}. Under similar conditions, the Unruh vacuum also appears as a thermal state on $i^+$ but differs from the Hartle--Hawking vacuum away from $i^+$ \citepbjps{dappiaggi2011rigorous}. The Unruh vacuum is defined to be a ground state for null out-going observers on the past horizon $\mathcal H^-$ of maximally-extended Schwarzschild and for static observers near $\mathscr I^-$. It is better physically motivated than the Hartle--Hawking vacuum as a description of the vacuum state of a collapsed black hole since any state evolving on a background that becomes Schwarzschild at late times will approach the Unruh vacuum near $i^+$, provided that the state satisfies certain regularity conditions in the ultraviolet limit and has no in-coming radiation from near spatial infinity $i^0$ (\citealtbjps{fredenhagen:1990}; \citealtbjps{hollands:2015}). In this sense, the Hawking spectrum is seen to be insensitive to the details of the collapse process and to the initial state of $\hat\phi$ on $\mathscr I^-$ (but away from $i^0$). The main constraint on the state can be formulated, for example, in terms of the Hadamard condition \citepbjps{hollands:2015} or via a scaling limit in the ultraviolet \citepbjps{fredenhagen:1990}. These constraints amount to regularity conditions in the ultraviolet limit and can be understood intuitively as enforcing high-energy modes to be effectively in a ground state. This avoids non-local divergences in the two-point functions (\citealtbjps{wald1994quantum}; \citealtbjps{hollands:2015}). Furthermore,  in conventional effective field theory treatments \citepbjps{candelas1980vacuum}, which assume no unusual effects in the ultraviolet due to quantum gravity, violation of the regularity conditions can be shown to lead to pathological behaviour near the horizon. 

To better understand the role of the regularity conditions in establishing the robustness of the thermal spectrum we can analyse the relative frequency shift between the Killing frequency and affine frequency of a particular Hawking mode as seen respectively by static and null-geodesic observers \citepbjps{Jacobson:1996}. The Killing frequency, which is constant along geodesics, is particularly useful for describing the frequency of free-falling modes of $\hat\phi$ as they approach future time-like infinity $i^+$, where the static observers become inertial. The affine frequency is suitable for describing the frequency of free-falling $\hat\phi$ modes near the horizon, where the affine coordinates describe a local patch of Minkowski space. A key formal property of the Schwarzschild geometry, which can be generalized to arbitrary Killing horizons \citepbjps{KAY199149}, is that the affine frequency is exponentially related to the Killing frequency near the horizon. Given the properties outlined above, this fact implies that free-falling modes near the horizon will be red-shifted exponentially as they approach $i^+$. Furthermore, if the Killing modes of $\hat\phi$ are expanded in terms of in-going and out-going affine components, then as the horizon is approached more and more of the in-going affine modes will disappear behind the horizon, which provides a causal barrier to their escape. Thus a static detector approaching $i^+$ will observe both an exponentially red-shifted spectrum and one entirely dominated by the out-going components. This means that the extreme red-shift combined with the presence of the horizon erases all information about how the horizon itself was formed and the details of the initial state of the radiative mode. The extreme red-shift further implies that in studying a moderate frequency mode at late times a static observer is effectively probing the ultraviolet structure of the two-point functions of the mode near the horizon. 

It is in this context that the regularity conditions imposed on the two-point function in the ultraviolet become vitally important. A noted above, the conditions require that the modes observed near $i^+$ be in their ground state near the horizon. Locally, this ground state is approximately the vacuum of Minkowski space split by the horizon into right and left Rindler-like regions. Tracing out the in-going modes (which disappear behind the horizon) therefore leaves the characteristically thermal state expected of the Rindler vacuum. 

The robustness of the thermal spectrum detected by late-time observers thus results from three ingredients:
\begin{enumerate}
\item Exponential red-shift (which implies that late-time observers probe ultra-high frequency modes near the horizon). 
\item Regularity conditions (which enforce that a state in the ultraviolet is approximately a Minkowski vacuum state). 
\item Event horizon (which creates a causal barrier between the in-going modes and the observers confined to the exterior region and therefore modifies the character of the state of the ultraviolet modes from `Minkowski-like' to `Rindler-like').
\end{enumerate}

\subsection{Ultraviolet Catastrophe} 

With some irony, the mechanisms that are responsible for the `remarkable robustness' of Hawking radiation are simultaneously a cause for scepticism.  As we have just seen, the exponential red-shift implies that the spectrum of radiation at late times is dominated by the characteristics of the state of the field near the horizon at energies that well exceed the Plank scale. In fact, any modes measured as cis-Planckian by stationary observers near $i^+$, must have originated as trans-Planckian modes from the point of view of free-falling observers less than a Planck unit of proper time before falling through the horizon. The Hawking radiation incident on a finite, stationary detector far away from the black hole can therefore be traced back to what are, for free-falling observers, trans-Planckian energies at the horizon. The above formulation is due to \citealtbjps{helfer:2003} and demonstrates that this `trans-Planckian problem' can be stated in an observer-independent way and thus cannot be ameliorated by redescription in a more fortunate coordinate system.\footnote{We will return to this point explicitly in the context of Polchinski's  `nice slice' argument shortly.} The energy of the near-horizon modes in question is in a regime where quantum gravity effects, such as those due to entanglement with the horizon, would be expected to be relevant. Thus, the implication of the trans-Planckian problem is that the semi-classical framework is not straightforwardly valid when applied to any effect, such as Hawking radiation, that is taken to be sensitive to near horizon physics.\footnote{The key historical papers on the trans-Planckian problem are (\citealtbjps{gibbons:1977}; \citealtbjps{unruh:1981}; \citealtbjps{Jacobson:1991}; \citealtbjps{Jacobson:1993}; \citealtbjps{unruh:1995}; \citealtbjps{PhysRevD.52.4559}). Accessible introductions are give in \citepbjps[\S7]{jacobson:2005} and \citepbjps[pp.36-8]{harlow:2016}. For discussion of the trans-Planckian problem in cosmology see \citepbjps{Martin:2001}.}  

As was aptly put by \citealtbjps{jacobson:2005}, the trans-Planckian problem amounts to `a breakdown in the usual separation of scales invoked in the application of effective field theory' (p.79).\footnote{For a more general argument that effective field theory methods may breakdown near horizons see \citepbjps{burgess:2018}.} In response to this breakdown, and in particular the role played by back-reaction and the evaporation process, \citealtbjps{fredenhagen:1990} conclude that `a full understanding of the [Hawking] phenomenon including a self-consistent description of the causal structure needs some elements of a quantum theory of gravity' (pp. 282--3). As noted by \citealtbjps{unruh2014has}, in typically robust fashion, `if one examines Hawking's original calculation, there are some severe problems with his derivation. While mathematically unimpeachable, they are nonsense physically' (p. 533). To highlight the significance of the scales involved Unruh estimates that the `frequencies which are needed to explain the radiation produced even one second after a solar mass black hole forms, correspond to energies which are $e^{10^5}$ times the energy of the whole universe.' \citepbjps[p. 533]{unruh2014has}. Plausibly, such considerations throw into doubt the idealizations and approximations upon which the entire semi-classical model of black holes is built.

An influential attempt to resolve the trans-Planckian problem is the `nice slice' argument due to  \citealtbjps{polchinski:1995} (see also \citealtbjps{Wallace:2017a}). Polchinski defines a particular slicing of a black hole spacetime that is `nice' in the sense that the slices are smooth, have small extrinsic curvature, and are such that in-falling particles are seen to have modest velocities. Due to the small extrinsic curvature the geometry of the slices changes slowly from slice to slice. Polchinski then argues that the adiabatic theorem should apply to modes in this slice and therefore that only very low-energy degrees of freedom can be excited from their ground state in the Hawking emission process (by whatever mechanism it takes place). According to this argument, the entire process can be described using local low-energy physics near the horizon and will therefore be independent of Planck scale effects. 

The `nice slice' argument is unconvincing as a response to the trans-Planckian problem as we have formulated it. In particular, the assumption that the relevant modes are in a genuine ground state is precisely the assumption that is in question. As emphasised by Jacobson in the quote above, the trans-Planckian problem occurs precisely because we have good reasons to expect the separation of energy scales to breakdown. Thus, whether the adiabatic theorem, or local arguments from effective field theory in general, should apply to near-horizon modes, which may be entangled or even interacting with the geometric degrees of freedom of the horizon, is exactly the issue at hand. Plausibly, some form of non-linear gravity--matter interaction is required for the evaporation process to take place at all. Furthermore, as noted by \citealtbjps{harlow:2016}, the adiabatic theorem applies only to the global conserved energy not to the centre of mass energy of localized excitations. Finally, the nice slice argument does nothing to ameliorate the exponential redshift: `it does not get rid of the fact that projecting onto possible final states of the late-time Hawking radiation produces states with a genuine high energy collision in the past' \citepbjps[pp. 37--8]{harlow:2016}. Thus, \textit{pace} \citealtbjps{Wallace:2017a}, Polchinski's argument offers no definitive means to rebut the force of the trans-Planckian problem.\footnote{The possibility for quantum gravity effects to undermine the nice slice argument is noted in \cite[\S2]{polchinski:1995} and in subsequent discussions regarding firewalls \cite{Almheiri:2012rt}.}

A different response to the trans-Planckian problem is based upon the  inconsistency of Hawking radiation not existing. The argument is that  the physics of the Planck scale cannot alter the Hawking spectrum since to do so would violate the semi-classical field equations (\citealtbjps{candelas1980vacuum};  \citealtbjps{sciama:1981}). However, once more, such a line of response is based upon an assumption that is itself in question. We might reasonably assume that due to the relatively mild curvature, the semi-classical field equations must apply to any local observers description of vacuum fluctuations near the horizon of an astrophysical black hole. However, in order for us to extend this assumption to near-horizon fluctuations as seen by a late-time observer, due to the exponential red-shift, we must assume that the semi-classical field equations continue to hold up to and beyond the Planck scale. However, it is at precisely at the Planck scale that we expect violations of the semi-classical field equations to occur. 
 
What we take to be the essential lesson is that absent a well-trusted theory of quantum gravity, any derivation of Hawking radiation as a phenomena that depends on near-horizon physics must be supplemented with an argument for the insensitivity of the effect to ultra short-distance physics.\footnote{To keep our discussion within reasonable constraints we have chosen to excluded approaches to Hawking radiation that do not feature near horizon sensitivity. See (\citealtbjps{parentani:2010}; \citealtbjps{giddings:2016}; \citealtbjps{hod:2016}; \citealtbjps{dey:2017}).} This supplement may take the form of an additional argument or a modification of the derivation. In either case, the insensitivity is required to both token-level and type-level. The first, 
 since we need to establish the insensitivity of the effect to different possible ultraviolet physics for a given type of black hole. The second, since to justify the idealizations (such as no backreaction) which the trans-Planckian problem gives us prima facie cause to doubt, we need to establish the insensitivity of the effect to the type of black hole being considered (e.g. astrophysical versus eternal).\footnote{As shall be discussed in Section 4.3, a further degree of type-level insensitivity relates to derivations of Hawking radiation that also apply to condensed matter analogue  black holes.} 

In the terminology introduced above what we are looking for is an argument for both the `robustness' and `universality' of Hawking radiation. It would be unreasonably restrictive to take the precise form of the radiative spectrum  to be insensitive to such token-level and type-level modifications. Rather, what we take to be the necessary, and arguably defining, property of the Hawking effect is the thermality of the detected radiation (at least to first order in $\hbar$). A token-level or type-level variation that destroys the thermality whilst preserving the particle flux is then a variation that undermines the Hawking effect (not least since it would allow the radiation to carry information). 

\section{Universality Arguments in Condensed Matter Systems}
\label{sec:Wilson}

\subsection{The Wilsonian Approach to Critical Phenomena}

A phase transition occurs when there is an abrupt change in the macroscopic parameters that uniquely specify the equilibrium states of a system. A first-order phase transition is characterized by the existence of discontinuities in the first derivatives of the free energy. Continuous phase transitions, in contrast, involve divergence of the response functions. An important feature of continuous phase transitions is that in the vicinity of the critical point measurable quantities depend upon one another in a power-law fashion. For example, in the ferromagnetic-paramagnetic transition, the net magnetization, $M$, the magnetic susceptibility, $\chi$, and the specific heat, $C$, all depend upon the reduced temperature $t= \frac{T-T_c}{T_c}$ (the temperature of the system with respect to the critical temperature $T_c$) as: 
\begin{equation}  
M\sim{|t|^{\beta}}, C\sim{|t|^{-\alpha}}, \chi \sim{|t|^{-\gamma}},  
\end{equation}
where $ \beta,$ $\alpha,$ $\gamma $ are the critical exponents. Another remarkable feature of these phenomena is the existence of cooperative behaviour at the transition or critical point, which means that the correlations between particles extend to very large distances even if the microscopic interactions between them remain short range. This implies the divergence of the correlation length $\xi$, which is a quantity that measures, for example, the distance over which spins in are correlated. The divergence of the correlation length is perhaps the most important feature of continuous phase transitions. In particular, it involves the loss of a characteristic scale at the transition point and thus provides a basis for the explanation of universal behaviour. That is, the divergence of the correlation length explains why system types with  physically  distinct  micro-structure, such as ferromagnets, antiferromagnets and fluids, display the same macro-behaviour. 

\citeposbjps{landau:1936} theory of continuous phase transitions  was one of the first attempts to give a rigorous explanation for the behaviour of physical variables close to the critical point and anticipated the development of renormalization group approaches.  In this theory, physical variables, such as magnetization, are replaced by their average values and non-linear fluctuation contributions are neglected. One can then use Landau theory to estimate the importance of fluctuations close to the critical point. For space dimension $d > 4$, the theory makes adequate predictions of the order parameter and the critical exponents.   Unfortunately, for $d \leqslant 4$, the Landau's theory of continuous phase transitions predicts strong infrared singularities in the lowest order fluctuation contributions, which means that fluctuations dominate the behaviour close to the critical point \citepbjps[\S 6]{Goldenfeld1992}. 

In analogy with the trans-Planckian problem, the problem of infrared singularities requires us to find a means to screen observable quantities in our theory from a breakdown in the separation of scales. One strategy is to explicitly renormalize the ultraviolet divergences, which involves expressing the weight of fluctuation contributions (amplitudes), in terms of physical coupling constants without assuming any particular cut-off in the calculation. Wilson \citeyearbjps{wilson:1971,wilson:1974} suggested a different strategy, now called momentum shell renormalization group (RG), that consists in integrating out short-range wavelength modes up to a finite cut-off in momentum $\Lambda$.\footnote{Kadanoff's real-space renormalization approach has been neglected for reasons of space. Plausibly our analysis will apply \textit{mutatis mutandis}. See  (\citealtbjps{Fisher1998}; \citealtbjps{Goldenfeld1992}; \citealtbjps{Mainwood2006}; \citealtbjps{Franklin2017}).} In this approach, one starts by defining a field theory in which the degrees of freedom are represented in terms of Fourier modes, $S(q)$. The partition function is then expressed as an integral over the full range of Fourier components. Each field theory, as defined by a particular local Hamiltonian (or Lagrangian), will then be characterized by the set of coupling constants ${g_i}$ that measure the strength of the various interactions. The core idea of the RG approach is to examine how the coupling constants change as one varies the length scale of interest, which is achieved by changing the value of the cut-off $\Lambda$. For continuous phase transitions one is interested in the limit of large length scales thus, for such theories, one analyses the behaviour of the coupling constants as the length scale increases. 

This RG procedure involves the following three steps: First, one carries out the partition integral over all Fourier components $S(q)$ with wave vectors residing in the momentum shell $\Lambda/b  \leqslant |q|  \leqslant \Lambda$, where $b > 1$. This step effectively eliminates the short-wavelength modes and thus corresponds to a coarse graining. Second, one relabels the control parameters by performing a scale transformation: $x \to x' = x/b$ and $q \to q' = b q$. Third, one relabels the field degrees of freedom by performing a scale transformation:  

\begin{equation}
 S (x) \to S'(x') = b^{\zeta } S(x), \\ S(q) \to S'(q') = b ^{\zeta - d} S(q),  \end{equation}   

where $\zeta$ must be chosen so as to assure that the rescaled residual Hamiltonian recovers the original form.   The three steps result in an `effective' Hamiltonian, which has different values of coupling constants than the original Hamiltonian but in successful cases has the same form. The RG transformations, given by repeated application of the three steps, are associated with a flow on theory space that  tells us how the coupling constants change. Depending on their behaviour under repeated iterations of the coarse graining transformations, the coupling constants can be classified as:  
\begin{enumerate}  
\item Relevant Coupling Constants: grow and ultimately tend to infinity as the number of iterations tends to infinity. These parameters allow one to define the critical surface;  
\item   Irrelevant Coupling Constants: ultimately approach zero in the RG procedure and do not affect the critical behaviour;  
\item  Marginal Coupling Constants: approach an infrared-stable fixed point that is associated to scale-invariant behaviour.    
\end{enumerate}  
The disappearance of the irrelevant couplings and the existence of non-trivial infrared fixed points is precisely what resolves the problem of infrared singularities in the RG approach. Crucially for our analysis, it is also this feature that establishes both the robustness and universality of the critical phenomena.   

The Wilsonian RG argument for universality takes the following form. Given the disappearance of the irrelevant couplings and the existence of a non-trivial infrared fixed point, the critical behaviour will depend only on the spatial dimension and the symmetries of the original Hamiltonians and not on the strength of the nonlinear couplings or other non-universal parameters. All distinct Hamiltonians whose trajectories converge toward the same infrared fixed point, i.e., the basin of attraction of the fixed point, will then exhibit identical behaviour at the critical point. This establishes universality since the critical phenomena in question have been shown to be insensitive to an inter-type variation between systems described by the set of distinct Hamiltonians (i.e., variation across the relevant universality class).  Furthermore, the same arguments also establish robustness. This is because the critical phenomena in question have been shown to be insensitive to a variation between possible microphysical realisations of a single type of system as described by distinct Hamiltonians. It is thus the very same mathematical properties of the basin of attraction of the fixed point that establish both robustness and universality. 

\subsection{The Six Qualities}

Drawing upon the Wilsonian exemplar we can identify six key qualities of successful universality arguments. First, the degree of robustness, which is the range of single-type token-level variation across which the invariance of the  macroscopic phenomena can be established. A Wilsonian treatment of RG transformations comes with a well-defined set of restrictions regarding the type of possible micro-interactions that can be shown to be irrelevant, and these limitations restrict the degree of robustness that the arguments can establish. One should also keep in mind that the theory space contains only theories within the broad framework of quantum field theory. When we are dealing with discrete systems, this assumption can obscure the connection to the microphysics of the systems. Further restrictions are then imposed within the family of field theories. In particular, the interactions must be short range \citepbjps[p. 161]{wilson:1974} and expressible in terms of a convergent set of constants and differential operators. Finally, in constructing theory space one is required to make assumptions about the number of spacetime dimensions and the irrelevance of, as yet unknown, fundamental spacetime structure. Not least in assuming a smooth $3+1$ spacetime model we are implicitly also assuming that we can rule out macro-level effects that have their origin in an unknown number of compactified extra dimensions or even fundamental dimensional heterogeneity \citepbjps{Tauber2012} due for example to quantum modifications to the spectral dimension of spacetime.  That said, we have good reason to expect that the `separation of scales' should mean that even if spacetime is fundamentally dimensionally heterogeneous or matter is not described by a quantum field theory, these assumptions will not undermine the effectiveness of the RG approach. These limits on the degree of robustness are thus very mild when considered in the relevant context and we can conclude that the Wilsonian universality arguments have a high degree of robustness. 

The second quality of Wilsonian arguments regards the degree of universality. This is the range of inter-type variation across which the  invariance of the phenomena can be established. This is one of the core virtues of Wilsonian renormalization group methods. In particular, the Wilsonian approach provides a means to characterise both quantitive and qualitative aspects of the `universal phenomena' across a wide range of system types such as fluids, ferromagnets, and antiferromagnets (\citealtbjps{Goldenfeld1992}; \citealtbjps{Tauber2012}).

The third quality is physical plausibility, which is closely connected to the first two. Physical plausibility is the applicability of the robustness and universality arguments to de-idealized target systems. The Wilsonian framework does particularly well on this front. In applying a renormalization group transformation, one coarse-grains out physically irrelevant detailed information until one reaches a non-trivial fixed point, which implies that the phenomenon under investigation is robust with regard to different possible microphysics and different possible complicating factors, both micro and macro (\citealtbjps{batterman:2014}; \citealtbjps{Palacios2017}). Moreover, it is the same arguments for robustness (insensitivity of single-type token level variation) that explain why scientific idealized models can be used to describe the behaviour of real target systems. In fact, one can use renormalization group arguments to demonstrate, for example, that real fluids and ferromagnets are in the same universality class as the idealized two-dimensional Ising model, so that the details that distinguish idealized models from real systems do not matter. As it will be seen below, the degree of physical plausibility is related to integration, since a lack of integration can limit the degree of physical plausibility.

The fourth quality is comprehensiveness. This is the size of the set of observables over which the robustness and universality argument can be applied. At criticality the arguments apply to all relevant observables and so this quality is again very strong. Although this quality is logically independent of the degree of robustness and universality, a high degree of comprehensiveness is required to have universality arguments that can successfully account for a rich class of phenomena. We will see in the next section that limiting the size of the set of observables over which universality arguments apply also limits the strength of these arguments. 

The fifth quality that characterises Wilsonian universality arguments is empirical support.  In fact, it is possible to use this framework to calculate explicit values of the critical exponents by linearizing around the fixed points and these values are found to be in good agreement with experimental results (see, for example, \citealtbjps{Ahlers}). Thus, the arguments are directly supported by experimental evidence in the relevant range of physically instantiated types. Furthermore, the methods of approximation that underlie the Wilsonian approach are also justified by the fact that they successful reproduce empirically observed phenomena.\footnote{See \citepbjps{blum:2016} for analysis of a historical case study focusing on the interplay between experimental evidence, approximations techniques and `emergent entities' in the context of RG techniques.} 

The sixth quality, crucial to our analysis, is integration. Integration is the quality whereby the theoretical bases behind the invariance found in the universality arguments and the robustness arguments are mutually consistent. Wilsonian arguments are clearly integrated since, as noted above, identical mathematical properties of the basin of attraction of the fixed point establish robustness and universality. Essentially the same arguments that successfully establish the irrelevance of the mico-details of a given system also allow us to explain the irrelevance of the details that distinguish different systems within the same universality class (see also Batterman \citeyearbjps{batterman:2002,batterman:2014}). Wilsonian universality arguments are constituted by the combination of the robustness argument together with general commonality conditions between the types based upon the spatial dimension and the symmetries of the original Hamiltonian. This means that a high degree of robustness will automatically imply a high degree of universality. Furthermore, arguments of this structure will also automatically be highly physically plausible. This is because we can use insensitivity under token-level variations between different possible micro-physics to show insensitivity under: i) de-idealization, which means that the same predictions will also hold for de-idealized realistic models; and ii) re-interpretation, which means that we can re-interpret the system type by adding details that characterise particular models.\footnote{It is important to note that this is not equivalent to the claim that one can find fixed-point solutions in a de-idealized model, which is still matter of controversy in the philosophical literature (\citealtbjps{batterman:2014}; \citealtbjps{Palacios2017}; \citealtbjps{saatsi:2018}).} 

 A different example that serves to illustrate the importance of integration for the physical plausibility of scientific models is the Fisher model for the explanation of the 1:1 sex ratio in diverse biological systems \citepbjps{batterman:2014}. In this case, it is the demonstration that a large class of details of particular systems (e.g. the actual population size) are irrelevant for the 1:1 sex ratio (i.e. robustness) that allows us to delimit the universality class of systems that will display the same behaviour. Moreover, since the Fisher model and real systems are in the same universality class, this also serves to demonstrate the physical plausibility of the model and to understand how such idealized model can be used to explain the 1:1 ratio in real biological populations.  As the Wilsonian case, we have integration since essentially the same arguments that successfully establish the irrelevance of the micro-details of a given system also allow us to explain the irrelevance of the details that distinguish different systems within the same universality class. In both cases, if the arguments were not integrated, and thus the universality and robustness arguments were inconsistent, physical plausibility would be severely limited. 

In general terms, integration is clearly a highly desirable feature since without integration only a limited amount of token-level insensitivity, as established by the robustness arguments, can be extendable to a different system types, via the universality arguments. When these types include physically realistic systems such lack of integration is also negatively related to physical plausibility since inconsistency between the robustness and universality arguments will imply both that token-level insensitivity is non-universal (i.e. highly type-specific), and that type-level insensitivity is non-robust (i.e. highly token-specific). We thus see that unintegrated arguments are likely to have limited physically plausibility and, moreover, not to be very useful in providing a general means to deal with problems due to a breakdown in the separation of scales. Unfortunately, it is precisely a lack of integration that we fill find in the context of the three sets of universality arguments for Hawking radiation considered in the next section.

\section{Universality Arguments for Hawking Radiation} 

\label{sec:universality_arguments_for_hawking_radiation}

\subsection{Arguments from the Equivalence Principle} 
\label{sec:unruh_universality_and_the_equivalence_principle}

Our first example is best understood as an independent universality argument for Hawking radiation, rather than an alternative derivation.  It is based upon a heuristic argument  that runs as follows. Through the equivalence principle, Einstein taught us that gravitation and acceleration are locally indistinguishable. This means that we should expect to be able to translate the local physics experience by a stationary observer in a region outside the event horizon of a black hole spacetime into equivalent physics experienced by a constantly accelerating Rindler observer in a Minkowski spacetime. A translation scheme based upon the equivalence principle would instruct us to identify the Hawking radiation detected by some stationary observer in the black hole spacetime with the Unruh radiation detected by a Rindler observer with the same acceleration. In particular, a simple argument---see for example \citepbjps [pp. 23-4]{Wallace:2017a}---can be used to numerically identify the Hawking temperature of a black hole with the Unruh temperature of a near-horizon observer red-shifted to infinity.\footnote{The red-shifting here is important because it regularizes the proper acceleration of a stationary observer on the horizon which is formally divergent.} Next, one shows that the Unruh effect is suitably robust under some class of possible ultraviolet modifications (i.e. quantum gravity effects). Using the translation scheme for near-horizon observers one then infers that the Hawking effect should also be robust to ultraviolet modifications. Such reasoning suggests that we were wrong to ever think of Hawking radiation and the Unruh effect as two separate phenomena, just as it would be wrong to distinguish gravitational and inertial mass. Rather, the equivalence  principle argument implies the two effects to be instances of single universal phenomenon of Unruh/Hawing radiation connected to acceleration/gravitation. Although suggestive, and to an extent physically insightful, in this section we will isolate the various senses in which such a line of argument is unreliable. We start by analysing the evidence for the robustness of the Unruh effect.  

The most physically salient starting point is the `detector approach' to describing Unruh radiation (\citealtbjps{unruh1976notes}; \citealtbjps{Unruh:1975gz}; \citealtbjps{Crispino:2008}). In this approach, a particle detector is made to follow a constantly accelerating path through Minkowski spacetime. In the simplest case, a scalar field in its vacuum state is coupled to the detector in such a way that the detector will count any sufficiently localized particle that enters the detector. The quantum field theoretic calculation can be done rigorously. In particular, the thermal spectrum can be obtained as the result of a particular branch cut in the relevant integrals over the divergent part of the two-point functions. Two important factors therefore determine the thermal form of this spectrum: the Lorentz invariance of the procedure, which controls the periodicity of the domain of the analytically continued integrals, and the precise form of the ultraviolet divergences of the two-point functions \citepbjps{Agullo:2009wt}. The physical mechanism behind the particle production in the Unruh effect can be understood in terms of the physical force responsible for the detector's sustained acceleration. The `energy reservoir' for the pair production is then seen to arise from whatever source is producing the energy to maintain this force. 

Given these insights, it is possible to investigate whether trans-Planckian modifications due to gravitational physics should have a noticeable effect on the spectrum observed by the detector via its response function. In this context, different kinds of ultraviolet modifications have been considered in the literature and all point to a limited degree of robustness of the thermal Unruh spectrum. In \citepbjps{Agullo:2009wt} the effect of introducing a Lorentz-invariant ultraviolet cutoff for the scalar field modes on the detector response function is considered. It is found that the thermal Unruh spectrum is insensitive to the introduction of this particular cutoff provided it is Lorentz-invariant. Contrastingly, it is also shown that non-Lorentz invariant cutoffs ruin the thermal properties of the spectrum by introducing a time-dependent damping effect on the detection rate of Unruh modes. We thus see that the Unurh effect should not be expected to be robust with respect to effects due to violations of Lorentz invariance in the ultraviolet. In more general terms, it can be explicitly shown via non-perturbative Functional Renormalization Group methods that, even in the case of Unruh radiation, the thermality of the spectrum can be ruined by a variety of quantum gravity effects \citepbjps{Alkofer:2016utc}. Such effects are unlikely to improve in the Hawking case, and could even be compounded by non-linear effects during the evaporation process. 

Certain features of Unruh radiation can thus be seen to be robust under certain limitations. On such a basis one might plausibly attempt to construct an argument for the universality of Hawking radiation based upon the equivalence principle along the lines of the heuristic argument presented above. \citepbjps{Agullo:2009wt} do this as follows. First, reason that a small detector near the horizon is locally indistinguishable from a Rindler observer in Minkowski space provided the detector is much smaller than the Schwarzschild radius of the black hole. Next, assert that such a detector will have a response rate identical to the one computed in the Unruh case. Finally,  use the robustness of the Unruh effect to reason that the response function of the detector in the black hole spacetime will be invariant under arbitrary Lorentz-invariant modifications.

There are goods reasons to be skeptical regarding this line of reasoning. As we have seen already, one must be cautious about using a local argument concerning the behaviour of a detector near a horizon to make inferences regarding the behaviour of a late-time detector far away from the horizon. While particle detection is itself a local process, the parameters of the Hawking spectrum, such as its temperature, depend functionally on global properties of the spacetime, such as its Arnowitt-Deser-Misner (ADM) mass. This is ultimately because the Hawking spectrum is determined in terms of an integrated effect of the field over its entire history. Moreover, as argued in detail by \citealtbjps{Helfer:2010ye}, the cis-Planckian Hawking modes detected by a stationary detector near the horizon of a black hole can contribute only a negligible proportion of the total Hawking flux detected at late times. Thus, almost all the late-time Hawking fluxes must stem from thermal modes that are trans-Planckian for the near-horizon inertial observers. It is therefore clear that universality arguments for Unruh radiation are completely ill-suited to provide a response to the trans-Planckian problem. They can only show insensitivity of near horizon cis-Planckian thermal spectrum to unknown trans-Planckian physics. By construction, they are silent regarding the properties of (almost all of) the late-time Hawking radiation. 

Equivalence principle based universality arguments thus do badly on the qualities of both degree of robustness and physically plausibility. The idealization that the entire class of observers can be represented by near-horizon detectors limits the argument in terms of both the token-level insensitivity established and applicability to de-idealized token system. In fact, it can be shown that repeating the above argument using any observer other than one infinitesimally close to the horizon leads to the wrong answer \citepbjps{Singleton:2011vh} for the predicted radiation near $i^+$.\footnote{This is effect is most extreme for static observers near $i^+$ who see no Unruh radiation at all.} This puts into the question the consistency of the entire approach. The score on degree of universality is slightly better since not only are the arguments applicable to Rindler and Schwarzschild spacetimes, they can also plausibly be extended to various other types of black hole spacetime including those representing astrophysical black holes formed via collapse. Contrastingly comprehensiveness is low since the set of observables over which the robustness argument can be applied is only contains one member. Empirical support is completely lacking. Furthermore, and most problematically, the argument is worryingly unintegrated. The theoretical basis for universality (the equivalence principe) in tension with the theoretical basis for robustness (effective field theory arguments applied to the Unruh effect). Plausibly, it is precisely the lack of integration that renders this an argument with rather limited qualities.  

\subsection{Arguments from Horizon Symmetries} 
\label{sec:arguments_from_horizon_symmetries}

Our next candidate university argument is best understood as a new derivation of Hawking radiation as a universal phenomena based upon the symmetries of the black hole event horizon. The argument relies upon the cancellation of anomalies of the effective event horizon symmetries to establish robust properties of the resulting Goldstone bosons. These Goldstone bosons are then interpreted as Hawking fluxes which suggests that Hawking radiation is itself robust. 

A good starting point is to observe that in general there exists an expansion of the d'Alembertian (in terms of the radial tortoise coordinate) near a fairly general class of horizons within which the leading order term is conformal and the angular contributions can be integrated out. This means that a Klein--Gordon theory reduces to an effective $1+1$ conformal field theory near such a horizon (\citealtbjps{Birmingham:2001qa}; \citealtbjps{Carlip:2005}). These simplifications allow, under fairly general assumptions, for the near-horizon Klein--Gordon modes to be written as representations of a chiral Virasoro algebra whose quantization is well-known. In $1+1$ dimensions, there is a conformal anomaly that can be expressed in terms of the topological invariants of the horizon geometry. This quantum mechanically broken conformal symmetry can be seen to lead to the generation of Goldstone bosons that must be present to cancel the anomaly. It is these Goldstone bosons that are interpreted as Hawking fluxes. Two formal observations support this interpretation. First, a state-counting of the Goldstone bosons generated by such an anomaly can be found to exactly reproduce the Bekenstein--Hawking entropy that one would expect for a black hole \citepbjps{Carlip:2005}. Second, if one computes the emission rate of the Goldstone bosons from the horizon, the response rate at infinity reproduces a thermal Hawking spectrum (\citealtbjps{Banerjee2010}; \citealtbjps{PhysRevD.77.024018}).\footnote{See also (Iso \textit{et al.} \citeyearbjps{Iso:2006ut,Iso:2006wa}).}

Arguably, further justification is needed to identify the generation of Goldstone bosons from a broken conformal symmetry with genuine Hawking fluxes. However, given such an identification, the argument from horizon symmetries shows precisely why we should expect Hawking radiation to have a degree of robustness comparable to the Wilsonian approach. In particular, that the anomalies in the near-horizon field theory are related to topological invariants of the horizon geometry means they can be expected to be invariant under a wide range of possible micro-physics. This is because any ultraviolet modifications that preserve the relevant symmetries without disrupting the topological properties of the horizon will produce the same spectrum of Goldstone bosons. This approach thus allows us to demonstrate invariance of the Hawking spectrum under any ultraviolet modifications that are due to quantum gravity effects that preserve the horizon symmetries without disrupting the topological properties of the horizon. Such assumptions are natural in string theory where conformal symmetry of the $1+1$ dimensional string worldsheet plays a vital role in the expected ultraviolet finiteness of the theory. However, the assumption that the horizon is smooth and has a definite position---even when probed by arbitrarily high energy modes---requires further justification in the context of quantum gravity. It is not even clear that the notion of horizon itself, which is a globally defined notion that relies on the existence of a time-like singularity, will survive in full theory of quantum gravity.\footnote{See \citepbjps{dougherty:2019} and \citepbjps{Wallace:2017a} for philosophical discussion of the potential issues with global definitions of black hole horizons.} Moreover, it is certainly not clear that physically realistic horizons, such as the event horizons of astrophysical black holes, can indeed be effectively described using Virasoro methods, since there is no proof that the reduction to $1+1$ dimensions can adequately encode the physics of collapse. This notwithstanding, given we have justification of the identification of Goldstone bosons as Hawking fluxes, we can take arguments from Horizon symmetries to imply a fairly wide range of  token-level insensitivity of Hawking radiation within the idealized system type that they describe. The arguments are thus moderately robust. 

A further advantage of these methods is in terms of their comprehensiveness. That is, the size of the set of observables over which the robustness argument can be applied. In particular, it is known that, for conformal field theories in $1+1$, all observables (i.e., $n$-point functions) can be computed directly from the symmetry considerations of the Virasoro algebra. This means that the arguments apply to all measurable processes rather than a restricted class of observables. With regard to the quality of comprehensiveness the horizon symmetry approach to black holes is comparable to the Wilsonian approach to condensed matter.

To what extent can we think of these arguments for horizon symmetries as universality arguments? Since only quite general properties of the horizon geometry and the Klein--Gordon operator are required to derive the Hawking flux, one might expect that the arguments will be applicable regardless of the type of physical system in question. Clearly this depends upon the extent to which the relevant features of the horizon geometry and the conformal expansion of the Klein--Gordon operator are instantiated in a wide range of physically plausible models. At present the Goldstone bosons and their entropies have only been computed for the a restricted class of horizons where back-reaction and evaporation effects are ignored. As noted above, such models are highly idealized. Most significantly, they assume both that the horizon can be treated as a boundary with fixed location even for the highest of trans-Planckian modes and that the equilibrium state of these modes is independent of the details of the non-stationary collapse process itself. It is an interesting open question whether the numerical coincidence between the Goldstone bosons and Hawking modes can be given a plausible physical basis in non-stationary and evaporating black hole models. However, it is certainly not a matter beyond all conjecture; if Hawking radiation really does causally depend upon the creation of the horizon and the violation of time translation invariance, perhaps due to some non-linear quantum back-reaction between the background quantum geometry and the trans-Planckian modes (or some other effect), then the physical basis behind this approach will be undermined. We thus find that both the physical plausibility and degree of universality of these arguments is severely limited. Empirical support is obviously also lacking.
 
Finally, we find arguments for the universality of Hawking radiation based upon horizon symmetries are not integrated. Unlike in the Wilsonian approach, the robustness arguments are highly type sensitive and thus inconsistent with relevant inter-type invariance arguments. It is natural to diagnose the weakness with regards to degree of universality and physical plausibly, despite high robustness in terms of a lack of integration. Thus, although arguments from horizon symmetries do provide a remarkably robust and comprehensive derivation of Hawking radiation, that establishes the effect as originating from very general features of anomaly cancellation, since they are unintegrated, these arguments do not establish the effect as universal and also there are doubts regarding their physically plausibility.  

\subsection{Arguments from Modified Dispersion Relations} 
\label{sub:arguments_that_use_modified_dispersion_relations}

Our final example of a universality argument for Hawking radiation takes the form of a general strategy for modifying derivations of Hawking radiation such that ultra-short distance effects are factored in. The key idea is that quantum gravity corrections to the Hawking spectrum can be modelled in terms of their effects on the propagation of the scalar field. In particular, the corrections are characterised in terms of a set of possible modifications to the dispersion relation of the high-energy Hawking modes. The late-time flux of Hawking modes is then computed with the modified dispersion relations using a straightforward generalisation of Hawking's original derivation. Provided the modifications to the dispersion relation satisfy a number of criteria the Hawking spectrum can be shown to be insensitive to the modifications.

The original idea (Jacobson \citeyearbjps{Jacobson:1991,Jacobson:1993};  \citealtbjps{unruh:1995}) behind the modified dispersion relation approach comes from numerical studies of analogue black holes,\footnote{See (\citealtbjps{unruh:1981}; \citealtbjps{garay:2000}; \citealtbjps{philbin:2008}; \citealtbjps{belgiorno:2010}; \citealtbjps{unruh:2012}; \citealtbjps{liberati:2012}; \citealtbjps{nguyen:2015}; \citealtbjps{Jacquet:2018}).} which indicate that one can use a modified dispersion relation to understand the `ultraviolet' breakdown of continuous fluid models due to atomic effects. Various generalisations of this approach to the gravitational case have now been achieved but, for our purposes, the most physically enlightening will prove to be that of \citealtbjps{unruh:2005}.\footnote{See also (\citealtbjps{PhysRevD.52.4559}; \citealtbjps{corley:1998}; \citealtbjps{himemoto:2000}; \citealtbjps{Barcelo:2008qe}; \citealtbjps{coutant:2012} \citealtbjps{Schutzhold:2013mba}).} 

The universality argument of \citealtbjps{unruh:2005} can be reformulated to make the comparison with Wilsonian universality arguments as follows.\footnote{Two particular differences in our formulation are that in their analysis condition ii is left implicit and condition iv is reformulated in a mathematically equivalent way (see its use in Equation~(16) of \citealtbjps{unruh:2005}).} First, take the family of modified dispersion relations to be parametrized by a single function, $F$, on momentum space. This function can be interpreted as representing a (non-Lorentz invariant) 3-momentum-dependent mass term for the Klein--Gordon field. In terms of the function $F$, the modified Klein--Gordon equation then takes the local form:
\begin{equation}\label{eq:mod dr}
	\left(\Box + F(\hat p^2)\right) \phi = 0\,,
\end{equation}
where $\Box$ is the Klein--Gordon operator in the exterior Schwarzschild spacetime and $\hat p^2$ is some differential representation of the square of the $3$-momentum operator in a first quantized Klein--Gordon theory. In momentum space, $F$ can be taken to be some function of the eigenvalue $k^2$ of the square of the linear momentum operator satisfying four criteria:

\begin{enumerate}
	\item [i.] Analyticity: $F(s^2)$ has an analytic continuation in terms of some convergent power series expansion in $s$, for $s \in \mathbb C$.
	\item [ii.] \label{vanishing infrared} Vanishing in the infrared: $F(k^2) \to 0$ when $k \to 0$ so that the dispersion relation is unmodified for the low energy modes.\footnote{A rapidity of convergence condition on could also be applied here.} 
	\item [iii.] Sub-luminal: $F(k^2) \leq k^2$ so that all modes travel slower than the speed of light.
	\item [iv.] \label{scaling limit} Scaling limit: $F(k^2) \to \tilde F_\infty k^2$ for $0< \tilde F_\infty < 1$ when $k\to \infty$ so that the dimensionless mass term, $\tilde F(k^2) = \frac {F(k^2)}{k^2}$, flows to a constant, $\tilde F_\infty$ in the ultraviolet.
\end{enumerate}
The relevant limits of $k$ above are defined relative to the Planck scale so that the infrared and ultraviolet limits represent the sub- and trans-Planck limits respectively. It is worth noting that, aside from the sub-luminal assumption, the above criteria are remarkably similar to those found in the Wilsonian renormalization framework. In particular, plausibly we can understand $F(k^2)$ as representing a running coupling in a particularly simple truncation of the Klein--Gordon theory space. The reasoning of this approach is to assume that $F(k^2)$ could be obtained from an honest Wilsonian treatment of the coupled Klein--Gordon and quantum gravity system. But without a concrete proposal for quantum gravity, the specific form of $F(k^2)$ is left free (up to the restrictions mentioned above). Moreover, in the context of the Hawking set-up where the Wentzel-Kramers-Brillouin (WKB) approximation applies to the Hawking modes, the classical (effective) Green's function contains most of the information of the full quantum mechanical two-point function. Thus criterion iv formally accomplishes many of the same things as requiring the Klein--Gordon state to be Hadamard or to have a scaling limit.\footnote{For the former, criterion iv in equation~(16) of \citealtbjps{unruh:2005} guarantees that all divergences of the Green's function are local. For the latter, the fixed point requirement of $\tilde F(k^2)$ is just equivalent to a scaling limit.} Using \eqref{eq:mod dr} it is possible to explicitly compute the spectrum of Hawking fluxes by essentially following Hawking's original procedure. The exponential nature of the red-shift between near-horizon Killing and affine modes drowns out any contributions coming from the power-series expansion of $F(k^2)$, and a thermal spectrum is straightforwardly recovered. 

The great strength of the modified dispersion relation approach is its physical plausibility. In fact, we take this style of approach to be the most physically plausible derivation of Hawking radiation available since, like the original Hawking calculation, we incorporate core physical features of astrophysical black holes whilst attempting to address certain aspects of the trans-Planckian problem. Since they are adapted from Hawking's original treatment of astrophysical black holes, modified dispersion relation approaches are embedded in a physically plausible context for Hawking radiation. Furthermore, as could be expected given its origin, the modified-dispersion approach is readily applicable to a huge range of physical systems and thus also has a high degree of universality. In fact, any system to which a Hawking-style derivation can be applied can be supplemented with a corresponding modified dispersion relation treatment. This includes various types of eternal and astrophysical black holes, and also a wide variety of analogue black hole systems. 

With regard to empirical support there is increasing cause for confidence. There are a large number of potential analogue realisations of the Hawking effect compatible with the modified dispersion relation argument (\citealtbjps{rousseaux:2008}; \citealtbjps{philbin:2008}; \citealtbjps{belgiorno:2010}; \citealtbjps{unruh:2012}; \citealtbjps{liberati:2012}; \citealtbjps{nguyen:2015}; \citealtbjps{Jacquet:2018}).  Furthermore, in some such cases the physical reliability of the techniques has been evaluated experimentally (\citealtbjps{rousseaux:2008}; \citealtbjps{weinfurtner:2011}; \citealtbjps{steinhauer:2016}).\footnote{In this context, it has been argued (\citealtbjps{Dardashti:2015a}; \citealtbjps{Dardashti:2019}; \citealtbjps{Thebault:2019}) that we can make inductive inferences about black holes based upon universality arguments combined with such `analogue experiments'. The success such an argument would also imply a degree of empirical support for the modified dispersion relation arguments in the context of black holes. For other examples of analogue experiments combined with universality arguments see (\citealtbjps{thouless1989}; \citealtbjps{prufer:2018}; \citealtbjps{erne:2018}; \citealtbjps{eigen:2018}).} Although modified dispersion relation arguments are not supported by direct experimental evidence for the full range of physically instantiated types, the methods of approximation that underlie the modified dispersion relation approach are (at least in part) justified by the fact that they successful reproduce empirically observed phenomena in some system types.

What we gain in physical reasonableness and degree of universality it appears, unfortunately, we have lost in degree of robustness. In particular, whilst the modifications considered are consistent with known microphysics of the analogue systems, they are highly restrictive with regard to the class of possible ultraviolet modifications to the physics of black holes. It is reasonable to identify this weakness in terms of a lack of integration. In particular, the same features that imply that the modified dispersion relations argument can be applied in a wider range of physical situations are those that limit its robustness in the specific context of black hole physics. A particular concern is whether sub-luminal modified dispersion relations really are sufficient to model quantum gravity effects. Most significantly, the truncation used in the `Klein--Gordon theory space' is of a very special type. In particular, a $k$-dependent sub-luminal mass term with a scaling limit in the ultraviolet is only the simplest example of how an effective quantum geometry could affect the propagation of Hawking modes across spacetime. Many quantum gravity proposals exhibit some sort of dimensional reduction or enhancement or non-locality in the ultra-violet where even a smooth local parameter such as $k$ (let alone an analytic expansion in terms of it) may not be defined. The space of all such effects cannot plausibly be modelled in such a limited truncation.\footnote{This can be seen explicitly in the quantum gravitational analysis of the Unruh effect mentioned above in \citepbjps{Alkofer:2016utc}.} The arguments also do not do well on comprehensiveness since they only apply  to the occupation number of the Hawking modes. 

In summary, the modified dispersion relation approach can legitimately be treated as a universality argument since it has the three qualities of robustness, physically plausibility and universality to a non-trivial degree. The arguments also have some empirical support. Modified dispersion relation arguments are, however, highly limited in their comprehensiveness. Moreover, with regard to black holes in particular, the arguments have only a low degree of robustness since there is only a very limited degree to which a modified dispersion relations approach can show the gravitational Hawking effect to be robust. Our diagnosis of the cause of this issue is again a lack of integration: in the context of black holes, the robustness and universality arguments are only consistent under a limited token-level variation, due to the limited class of ultra-violet modifications that are consistent with the modified dispersion relation approach.

\section{Conclusion}
\label{sec:comparison_hawking_vs_wilson_universality}

Given that the trans-Planckian problem threatens to undermine the entire semi-classical framework for modelling black holes, it is significant that none of the available universality arguments offer an entirely convincing means of response to it. Universality arguments for Hawking radiation in the context of black hole physics do poorly in comparison with the Wilsonian arguments in the context of condensed matter physics on their own terms (that is, over and above more general epistemic worries relating to the quantum description of black hole physics).  A natural way forward drawn from the analysis above would be to attempt to combine the horizon symmetry and modified dispersion relation arguments for the universality of Hawking radiation. However, the coherence of such a combined approach is yet to be seen. In particular, the stationarity idealization used in the anomaly cancellation approach is not compatible with a Hawking-style derivation where the (non-stationary) process of the formation of the horizon plays an important role. There is, as yet, no smooth limit in which the original Hawking derivation can be seen to lead to the context in which the anomaly cancellation arguments are based. A significant challenge is to show that  anomaly cancellation arguments can be successfully applied to a more realistic model of a collapsing black hole. In particular, there is no existing universality argument or analogue model that deals explicitly with how quantum fluctuations and back-reaction of the horizon, which should dominate the physics of the trans-Planckian modes, could impact the thermality of the Hawking spectrum. If that could be achieved, then it is plausible that a suitably integrated universality argument for Hawking radiation could be established based upon the combination of the modified dispersion relation and horizon symmetry approaches.

\section*{Funding}

Work on this paper was supported by the Arts and Humanities Research Council, UK (Grant Ref. AH/P004415/1).

\section*{Acknowledgments}
This paper has benefited from discussions with a large number of different people over a period of years. We would like to give particular thanks to Vincent Ardourel, Alex Blum, Radin Dardashti, Stephan Hartmann, Maxime Jacquet, Ted Jacobson, S\'ebastien Rivat, Katie Robertson, Charlotte Werndl, Eric Winsberg, audience members at talks in Cambridge, Munich and Nijmegen, and two anonymous referees. Finally, we are deeply indebted to Erik Curiel and David Wallace for extensive dicussions and detailed comments on a draft manuscript. 

\begin{flushright}
\emph{
  Sean Gryb\\
  University of Groningen\\
  Faculty of Philosophy \\
  Groningen, The Netherlands\\
  s.b.gryb@rug.nl\\
}
\bigskip
\emph{
  Patricia Palacios\\
  University of Salzburg\\
  Department of Philosophy\\
  Salzburg, Austria\\
  patricia.palacios@sbg.ac.at
}\\
\bigskip
\emph{
  Karim Th\'ebault\\
  University of Bristol\\
  Department of Philosophy\\
  Bristol, UK\\
  karim.thebault@bristol.ac.uk
}
\end{flushright}

\bibliographystyle{bjps}
\bibliography{dumb}{}

\end{document}